\documentclass{jltp}

\usepackage{graphicx} 

\title{Landau and Dynamical Instabilities
of Bose-Einstein Condensates with Superfluid Flow
 in a Kronig-Penney Potential}

\author{Ippei Danshita$^{a,b}$ and Shunji Tsuchiya$^c$}

\address{$^a$National Institute of Standards and Technology, 
\\
Gaithersburg, MD 20899, USA
\\
$^b$Department of Physics, Waseda University, \=Okubo, Shinjuku, 
\\
Tokyo, 169-8555, Japan
\\
$^c$CNR-INFM BEC Center and Dipartimento di Fisica, Universit\`a di Trento,
\\
I-38050 Povo, Trento, Italy
}

\runninghead{I. Danshita and S. Tsuchiya}{Landau and dynamical instabilities of BECs in a KP potential}

\begin{document}

\maketitle

\begin{abstract}
We study the elementary excitations of Bose-Einstein condensates in a one-dimensional periodic potential and discuss the stability of superfluid flow based on the Kronig-Penney model. We analytically solve the Bogoliubov equations and calculate the excitation spectrum. 
The Landau and dynamical instabilities occur in the first condensate band when the superfluid velocity exceeds certain critical values, which agrees with the result of condensates in a sinusoidal potential. 
It is found that the onset of the Landau instability coincides with the point where the perfect transmission of low-energy excitations is forbidden, while the dynamical instability occurs when the effective mass is negative. 
It is well known that the condensate band has a peculiar structure called swallowtail when the periodic potential is shallow compared to the mean field energy. 
We find that the upper side of the swallowtail is dynamically unstable although the excitations have the linear dispersion reflecting the positive effective mass.

\end{abstract}

\section{INTRODUCTION}
The realization of Bose-Einstein condensates (BECs) in a optical lattice has attracted much attention of both theorists and experimentalists~\cite{rf:rev}.
One of the interesting problems in this system is the breakdown of the superfluidity.
In a recent experiment, the stability of BECs in a moving 1D optical lattice has been investigated, it has been demonstrated that the Landau and dynamical instabilities destabilize the system when the velocity of the lattice exceeds certain critical values~\cite{rf:moving}.
In addition, in a number of theoretical works, the stability of BECs in an optical lattice have been studied in great detail~\cite{rf:wuniu,rf:pesmith,rf:edo,rf:smerzi,rf:konabe,rf:iigaya}.

In the present paper, by means of the Kronig-Penney (KP) model, we study the elementary excitations and the stability of BECs with supercurrent in a 1D periodic potential.
In our previous work, the excitation spectrum of current-free BECs in a Kronig-Penney potential has been calculated by applying the Bloch theorem to the solution of Bogoliubov equations for a single-barrier problem~\cite{rf:wareware}.
We will extend this theory to the case of a current-carrying BECs.

\section{MEAN FIELD THEORY AND CONDENSATE BAND}
We consider a quasi-1D BEC at the absolute zero of temperature in a Kronig-Penney potential, which is a periodic array of $\delta$-function potential barriers,
  \begin{eqnarray}
  V(x)=V_0\sum_{j=-\infty}^{\infty}\delta(x-ja),
  \end{eqnarray}
where $a$ is the lattice constant, and $V_0$ is the potential strength.
We consider that the condensate has supercurrent flowing through the periodic potential, which corresponds to the situation where the optical lattice created by two counter-propagating laser beams with frequency difference is moving at a constant velocity~\cite{rf:moving}.

Our formulation of the problem is based on the mean field theory, which consists of the time-independent Gross-Pitaevskii equation and the Bogoliubov equations~\cite{rf:BEC}.
Assuming that the condensate is weakly perturbed and seeking the solution of the time-dependent Gross-Pitaevskii equation in the form $\Psi(x,t)=\Psi_0(x)+u(x){\rm e}^{-\frac{{\rm i}\varepsilon t}{\hbar}}-v(x)^{\ast}{\rm e}^{\frac{{\rm i}\varepsilon t}{\hbar}}$, one obtains them,
  \begin{eqnarray}
  \left[-\frac{\hbar^2}{2m}\frac{d^2}{d x^2}-\mu+V(x)
  +g|\Psi_0(x)|^2\right]\Psi_0(x) =0,\label{eq:sGPE}
  \end{eqnarray}
and
       \begin{eqnarray}
               \left(
                 \begin{array}{cc}
                 H_0 & -g{\Psi_0(x)}^2 \\
                 g{\Psi_0(x)}^{\ast2} & -H_0
                 \end{array}
               \right) 
               \left(
                 \begin{array}{cc}
                 u(x) \\ v(x)
                 \end{array}
               \right)
               = \varepsilon\left(
                 \begin{array}{cc}
                   u(x) \\ v(x)
                 \end{array}
               \right),  \label{eq:BdGE}\\
               H_0 = -\frac{\hbar^2}{2m}\frac{d^2}{d x^2}
               -\mu+V\left(x\right)+2g|\Psi_0|^2,
       \end{eqnarray}
where $\mu$ is the chemical potential and $g$ is the coupling constant of the interparticle interaction.
The excitation energy $\varepsilon$ tells us the stability of a BEC.
The appearance of the negative excitation energy is a signature of the Landau instability, while the appearance of the complex excitation energy is that of the dynamical instability.

\begin{figure}[b]
\begin{center}
\includegraphics[width=\linewidth, height=3 cm]{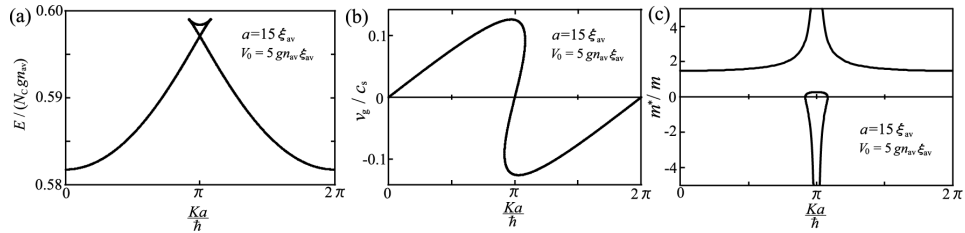}
\end{center}
\caption{
First condensate band (a), group velocity (b), and effective mass (c) with $(a,V_0)=(15 \xi_{\rm av},5 gn_{\rm av}\xi_{\rm av})$, where $\xi_{\rm av}\equiv\frac{\hbar}{\sqrt{mgn_{\rm av}}}$, $c_{\rm s}\equiv\sqrt{\frac{gn_{\rm av}}{m}}$, $n_{\rm av}\equiv\frac{N_{\rm C}}{a}$, and $N_{\rm C}$ is the number of condensate atoms.
} 
\label{fig:all}
\end{figure}

At first, we shall solve Eq. (\ref{eq:sGPE}).
We assume that the condensate stands in the first condensate band.
Then, assuming that the lattice constant $a$ is much larger than the healing length $\xi$, one can approximately obtain the solution of Eq. (\ref{eq:sGPE}) as
  \begin{eqnarray}
  \Psi_0(x)=\sqrt{\frac{\mu-3\epsilon_Q}{g}}
  \left({\rm tanh}\left(\frac{|x-ja|+x_0}{\xi}\right)
  -{\rm sgn}(x){\rm i}\frac{Q\xi}{\hbar}\right)
  {\rm e}^{{\rm i}\left(\frac{Qx}{\hbar}-{\rm sgn}(x)\theta_0\right)},
  \nonumber
  \end{eqnarray}
\vspace{-5mm}
  \begin{eqnarray}
  \left(j-\frac{1}{2}\right)a<x<\left(j+\frac{1}{2}\right)a,
  \label{eq:con}
  \end{eqnarray}
where $\epsilon_Q\equiv\frac{Q^2}{2m}$ and $\xi\equiv\frac{\hbar}{\sqrt{m(\mu-3\epsilon_Q)}}$.
Constants $x_0$ and $\theta_0$ are determined by the boundary conditions at $x=ja$.

Imposing the Bloch theorem $\Psi_0(x+a)=\Psi_0(x){\rm e}^{\frac{{\rm i}Ka}{\hbar}}$ on the condensate wave function (\ref{eq:con}), we can calculate the first condensate band $E$, the group velocity $v_{\rm g}\equiv\frac{\partial E}{\partial K}$ and the effective mass $m^{\ast}\equiv\left(\frac{\partial^2 E}{\partial K^2}\right)^{-1}$ as functions of the condensate quasimomentum $K$ as shown in Fig. \ref{fig:all}.
The swallowtail structure in the first condensate band exists around the edge of the first Brillouin zone (Fig. \ref{fig:all}(a)), which disappears when the potential strength is sufficiently large compared to the mean field energy~\cite{rf:seaman}.
Several authors have pointed out that the upper side of the swallowtail is dynamically unstable~\cite{rf:pesmith,rf:mueller,rf:seaman}.
On the other hand, the system is expected to be stable for the long wavelength phonon in the upper side, because the effective mass is positive (Fig. \ref{fig:all}(c)).
We will confirm these two fact, namely the dynamical instability and the existence of the stable phonons in the upper side.

\section{EXCITATION SPECTRUM AND STABILITY}
Next, we analytically calculate the excitation spectrum.
We shall first find solutions of the Bogoliubov equations (\ref{eq:BdGE}) in the region $|x|<\frac{a}{2}$ where only a single barrier exists.
There exists two independent solutions with the excitation energy $\varepsilon$ of the single barrier problem, corresponding to two types of scattering process.
One is the process in which an excitation comes from the left ($\psi^{\rm l}(x)$), and the other from the right ($\psi^{\rm r}(x)$).
$\psi^{\rm l}(x)$ is written as~\cite{rf:ware}
  \begin{eqnarray}
   \psi^{\rm l}(x)
   =
     \left(\!\begin{array}{cc}
                      u^{\rm l} \\ v^{\rm l}
                  \end{array}\!\right)
   \!\!\!&=&\!\!\!\left\{\begin{array}{ll}
          \left(\!\begin{array}{cc}
                    u_1 \\ v_1
                         \end{array}\!\right)
          +r^{\rm l}\left(\!\begin{array}{cc}
                    u_2 \\ v_2
                         \end{array}\!\right)
          +a^{\rm l}\left(\!\begin{array}{cc}
                    u_3 \\ v_3
                         \end{array}\!\right),
                         & x<0,  \\
          t^{\rm l}\left(\!\begin{array}{cc}
                    u_1 \\ v_1
                         \end{array}\!\right)
          +b^{\rm l}\left(\!\begin{array}{cc}
                    u_4 \\ v_4
                         \end{array}\!\right),
                         & x>0,
          \end{array}\right. \label{eq:lcs}
\end{eqnarray}
where the coefficients $r^{\rm l}$, $a^{\rm l}$, $t^{\rm l}$, and $b^{\rm l}$ are the amplitudes of the reflected, the left localized, the transmitted, and the right localized components, respectively.
All the coefficients can be determined by the boundary conditions at $x=0$.
Once the coefficients of $\psi^{\rm l}(x)$ is obtained, the coefficients $r^{\rm r}$, $a^{\rm r}$, $t^{\rm r}$, and $b^{\rm r}$ of $\psi^{\rm r}(x)$ can be easily obtained as well~\cite{rf:ware}.
The transmission coefficient $|t^{\rm l}|^2$ strongly depends on the supercurrent as follows.
If there is no supercurrent, the transmission coefficient has a kind of resonant behavior with a peak at $\varepsilon=0$, which is called {\it anomalous tunneling}~\cite{rf:antun}.
As the supercurrent increases, the width of the peak becomes narrow; consequently, the anomalous tunneling disappears at a certain critical value of the supercurrent~\cite{rf:ware}.

We can write a general solution of the Bogoliubov equations (\ref{eq:BdGE}) in $|x|<\frac{a}{2}$ as a linear combination of $\psi^{\rm l}(x)$ and $\psi^{\rm r}(x)$:
  \begin{eqnarray}
  \psi(x)=\alpha\psi^{\rm l}(x)+\beta\psi^{\rm r}(x),\,\,\,|x|<\frac{a}{2},
  \end{eqnarray}
where $\alpha$ and $\beta$ are arbitrary constants.
Extending this solution to all regions of $x$ by means of the Bloch theorem, one can obtain the relation between the energy $\varepsilon$ and the quasimomentum $q$ of the excitation:
  \begin{eqnarray}
  {\rm exp}\left[{\rm i}\frac{(-k_1+k_2)a}{2\hbar}\right]
  +(t^{\rm l}t^{\rm r}-r^{\rm l}r^{\rm r})
  {\rm exp}\left[{\rm i}\frac{(k_1-k_2)a}{2\hbar}\right]=
  \nonumber\\
  t^{\rm l}{\rm exp}\left[{\rm i}
  \frac{(-2q+k_1+k_2)a}{2\hbar}\right]
  +t^{\rm r}{\rm exp}\left[{\rm i}
  \frac{(2q-k_1-k_2)a}{2\hbar}\right].
  \label{eq:rel_eq}
  \end{eqnarray}
where $k_{1,2}$ satisfies the Bogoliubov excitation spectrum in a uniform system~\cite{rf:fetter},
  \begin{eqnarray}
  \varepsilon=\frac{Qk}{m}
  +\sqrt{\frac{k^2}{2m}\left(\frac{k^2}{2m}+2(\mu-\epsilon_Q)\right)}.
  \end{eqnarray}
As well as the case of a current-free condensate, we clearly see from Eq. (\ref{eq:rel_eq}) that the tunneling properties in the single barrier problem determine the band structure of the excitation spectrum.

\begin{figure}[tb]
\begin{center}
\includegraphics[width=\linewidth, height=4.0cm]{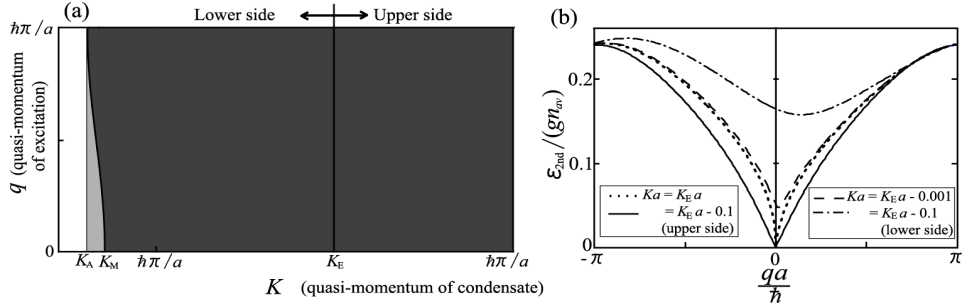}
\end{center}
\caption{\label{fig:stability}
(a) Stability phase diagram of the condensate in a KP potential, where $(a,V_0)=(15\xi_{\rm av},5gn_{\rm av}\xi_{\rm av})$.
The light shaded area and the dark shaded area correspond to the regions of the Landau instability and the dynamical instability, respectively.
This figure is focused to a region close to the swallowtail.
(b) Second band of the excitation spectrum.
}
\end{figure}
Now let us discuss the stability of condensates in a KP potential.
We note in advance that there are three important values of the condensate quasimomentum, $K_{\rm A}$, $K_{\rm M}$ and $K_{\rm E}$.
$K_{\rm A}$ corresponds to the point where the anomalous tunneling disappears, $K_{\rm M}$ to the point where the group velocity reaches its maximum value and the sign of the effective mass changes, and $K_{\rm E}$ to the edge of the swallowtail, respectively.

Calculating the excitation spectrum by solving Eq. (\ref{eq:rel_eq}), we obtain the stability phase diagram as shown in Fig. \ref{fig:stability}(a).
The light (dark) shaded area corresponds to the region of the Landau (dynamical) instability where the negative-energy (complex-energy) excitation exists.
Both of the instabilities appear in the first band of the excitation spectrum.
In the lower side of the swallowtail, $K=K_{\rm A}$ is the onset of the Landau instability around $q=0$ and the dynamical instability around $q=\frac{\hbar\pi}{a}$, while $K=K_{\rm M}$ is the onset of the dynamical instability around $q=0$.
In the upper side, the system is always dynamically unstable, which agree with the result of several recent works.
On the other hand, as shown in Fig. \ref{fig:stability}(b), there also exists the gapless and linear dispersion in the upper portion, reflecting the positive effective mass.
As $K$ in the lower portion of the swallowtail approaches $K_{\rm E}$, the bottom of the second band of the excitation spectrum approaches the origin. When $K$ reaches $K_{\rm E}$, the second band turns into a gapless dispersion.
Consequently, there exists the gapless and linear dispersion in the upper portion, reflecting the positive effective mass, even though the system is dynamically unstable due to the first Bogoliubov band.
\section{CONCLUSION}
In summary, we have investigated the excitation spectrum and the stability of BECs in a KP potential.
We have found that the onset of the Landau instability coincides with the point where the anomalous tunneling disappears, while all the excitations in the first band of the excitation spectrum exhibits the dynamical instability when the effective mass is negative.
It has been shown that the second band of the excitation spectrum takes the gapless and linear form for the long-wavelength excitations in the upper side of the swallowtail, while the system is always dynamically unstable.
\section*{ACKNOWLEDGMENTS}
The authors greatly acknowledge useful comments of D. L. Kovrizhin.
I.D. is supported by JSPS (Japan Society for the Promotion of Science) Research Fellowship for Young Scientists.

\end{document}